\documentclass[%
 reprint,
 amsmath,amssymb,
 aps,
]{revtex4-2}
\usepackage{draftwatermark} 
\usepackage{graphicx}
\usepackage{dcolumn}
\usepackage{bm}
\usepackage{pgfplots}
\usepackage{tabularx}
\usepackage{booktabs}
\usepackage{algorithmic}
\usepackage{siunitx}
\usepackage[ruled,linesnumbered]{algorithm2e}

\begin{document}

\preprint{APS/}

\title{Non-local transport in Radiation-Hydrodynamics codes for ICF by efficient coupling to an external Vlasov-Fokker-Planck code}

\author{Abetharan Antony} \email{abetharan03@gmail.com}
\author{Robert J Kingham}%
\affiliation{%
Blackett Lab., Plasma Physics Group, Imperial College London, London, SW7 2AZ, UK
}%
\author{Stefan Mijin}
\affiliation{
United Kingdom Atomic Energy Authority, Culham Campus, Abingdon, Oxfordshire, OX14 3DB, UK}

\author{Michael M Marinak}
\affiliation{Lawerence Livermore National Laboratory,7000 East Ave, Livermore, CA 94550, United States}%

\date{\today}

\begin{abstract}
Accurately incorporating non-local transport into radiation-hydrodynamics codes, and indeed any fluid system, has long been elusive. To date, a simplified and accurate theory that can be easily integrated has not been available. This limitation affects modeling in inertial confinement fusion and magnetic confinement fusion systems, among others, where non-local transport is well-known to be present. Here, we present a coupling methodology between a full Vlasov-Fokker-Planck (VFP) electron kinetic code and radiation-hydrodynamics (rad-hydro) codes. The VFP code is used to adjust native electron transport in the rad-hydro code, thus enabling improved transport without the need to integrate a full electron VFP solver into the rad-hydro code. This approach necessitates only occasional invocation of the VFP code, reducing computational intensity compared to following the dynamic evolution entirely with the VFP code on fluid time scales. We illustrate that the methodology is more accurate than other simplified methods in thermal decay systems relevant to inertial confinement fusion and can replicate standard theoretical results with high accuracy.
\end{abstract}

\maketitle

\SetWatermarkText{PREPRINT} 
\SetWatermarkScale{3} 
\section{\label{sec:level1} Introduction}
Inertial Confinement Fusion (ICF) and Magnetic Confinement Fusion (MCF) present numerous computational challenges within plasma physics. A primary limitation in numerical modeling arises from the approximations made to maintain simplicity or computational efficiency- especially in electron transport, a critical component in various fusion schemes and the focal point of this work. In Indirect-Drive ICF, for instance, a laser heats the inner walls of a high-Z cylinder—a hohlraum—creating X-rays that compress a fuel capsule. Electron transport affects energy deposition and redistribution in the hohlraum plasma, influencing both X-ray conversion efficiency and symmetry \cite{Atzeni2004}. In Direct-Drive ICF, a laser heats the fuel capsule directly, generating fast electrons that can preheat the cold fuel \cite{Hill2019}. In tokamaks with divertors, accurate modeling of parallel electron heat transport in the scrape-off layer is essential for managing heat loads on divertor plates \cite{Mijin2020Thesis}.

Electron transport has traditionally been approximated as small deviations from equilibrium behavior (classical transport) \cite{Braginskii1965, Spitzer1953}. However, in many fusion regimes, the electron distribution often deviates significantly from a Maxwellian distribution \cite{Rinderknecht2018}. The regime where electron kinetic transport becomes essential is defined by the Knudsen number, $K_n = \lambda_{mfp} / L$, where $\lambda_{mfp}$ is the electron mean free path, and $L$ is the characteristic scale length of the system. Classical transport is typically valid only when $K_n \ll 0.06/\sqrt{\Bar{Z}}$ \cite{Brantov2013}. Non-local behavior occurs in multiple fusion schemes, including the scrape-off layer (SOL) in MCF, hohlraums in Indirect ICF, and capsule ablation in Direct-Drive ICF.

In simulations, electron transport is often modeled using flux-limited variants of Braginskii \cite{Braginskii1965} or Spitzer-Harm \cite{Spitzer1953}, with flux limiters tuned experimentally or chosen \textit{ad hoc}. However, these models are not generalizable, and simulations frequently struggle to match experimental values. For instance, this mismatch may contribute to the so-called ‘drive deficiency’ in indirect-drive ICF, where simulations over-predict X-ray flux by 20-30\% compared to experimental results \cite{Moody2014}.

The SNB model \cite{Schurtz2000}, now standard in most ICF codes \cite{Brodrick2017, Marocchino2013}, also shows considerable discrepancies, particularly in three areas: (1) flux suppression, (2) pre-heating, and (3) distribution function accuracy \cite{Brodrick2017, Sherlock2017a}. The first two issues affect thermal conductivity, while distribution function accuracy impacts other transport coefficients, such as electrical resistivity and thermoelectric coefficients, both of which influence magnetic induction. These limitations underscore the need for improved transport modeling to capture non-local electron behavior accurately.

The predictive shortcomings of flux limiters and approximate models reveal a gap in current methodologies, particularly in capturing non-local effects in electron transport under extreme conditions necessary for ICF and MCF applications. These deficiencies may contribute to phenomena like the ‘drive deficiency’ in indirect-drive ICF. Addressing these limitations necessitates a novel approach, where coupling kinetic codes with radiation hydrodynamics presents a promising solution to model non-local electron behavior with improved fidelity and computational efficiency.



To fully capture the dynamics of electron transport, we start with the complete kinetic formulation for a plasma species, which describes the evolution of the distribution function  f(\textbf{r},\textbf{v},t)  in six-dimensional phase space:
\begin{equation}
\label{eq:vfp}
\frac{\partial f}{\partial t} + \nabla_{\textbf{r}}\cdot(\textbf{v}f,) + \nabla_{\textbf{v}} \cdot(\textbf{F}f,) = \bigg(\frac{\partial f,}{\partial t}\bigg)_{coll},
\end{equation}
where $f(\textbf{r},\textbf{v},t)$ is the distribution function in 6D phase space, and $\textbf{F}$ is the Lorentz force due to smoothed macroscopic fields. The RHS captures the effects of species collisions.. The Lorentz force is given by
\begin{equation}
\mathbf{F} = q(\mathbf{E} + \mathbf{v}\times\mathbf{B}),
\end{equation}
where $\mathbf{F}$ is the Lorentz force on a particle of charge $q$, moving with velocity $\mathbf{v}$ in an electric ($\mathbf{E}$) and magnetic field ($\mathbf{B}$). The two sides of this Vlasov-Fokker-Planck (VFP) equation capture distinct aspects of plasma behavior: the LHS is the Vlasov equation, which describes collective plasma behavior, while the RHS accounts for microscopic behavior due to collisions. Together, these represent what is commonly called a full $f$ description of plasma.

However, solving the full kinetic equation is impractical for both analytical and computational purposes. Therefore, it is common to expand the distribution function in velocity space using spherical harmonics as the basis function and truncate the expansion at a finite order. This method reduces the three-dimensional velocity dependence to one dimension, with the angular distribution function information captured by the spherical harmonic coefficients. This is a beneficial expansion as electron-ion scattering tends to isotropize the electron distribution function, causing higher-order ($l$) components to decay rapidly. This allows us to accurately represent the distribution function using a limited number of spherical harmonics which reduces computational expense. Resulting in the following expansion in one spatial dimension (see \cite{Shkarofsky1963} for more details):

\begin{equation}
f(v,\theta,\phi, \mathrm{t}) = \sum_{l=0}^{\infty}\sum_{m=-l}^{l} f^{m}_l(v, \mathrm{t}) P^m_l(\cos \theta)e^{i m \phi},
\end{equation}
where $f^{-m}_l = (f^m_l)^*$, with $\theta = 0$ in the $x$ direction and $\phi = 0$ in the $y$ direction. 

It is important to distinguish the physical meanings of the decomposition. The zeroth-order distribution function in the expansion, $f_0$, is related to density and energy as follows:
\begin{gather}
E = \frac{m_e}{2}\int_0^\infty v^4 f_0 \mathrm{d}v,\
n_e = \int_0^\infty v^2 f_0 \mathrm{d}v,
\end{gather}
while $f_1$ is related to vector quantities, such as heat flow, given by:
\begin{equation}
\mathbf{q} = \frac{m_e}{2}\int_0^\infty v^5 f_1 \mathrm{d}v.
\end{equation}

This formulation of plasma dynamics represents a high-fidelity model for transport. However, it is numerically expensive and challenging to incorporate essential physics — such as self-consistent radiation transport. To overcome these challenges, coupling kinetic and fluid models has emerged as an effective strategy for achieving accurate transport modeling while retaining the computational efficiency of fluid approaches.

While such coupling is common in Magnetic Confinement Fusion (MCF) studies \cite{Chankin_2022, Zhao2017, Zhao2019a, Mijin2020phys}, Inertial Confinement Fusion (ICF) simulations have traditionally relied on radiation hydrodynamics codes with electron heat fluxes capped by flux limiters, determined either phenomenologically or informed by kinetic simulations \cite{Bell1981}. Some efforts have introduced kinetic effects into ICF—for example, simulations of Direct-Drive experiments employing a cold ion fluid model alongside electron kinetics \cite{Sunohara2003, Hill2021, Joglekar2016} — but these approaches fall short for hohlraum modeling where radiation transport is critical. Recently, kinetic electrons have been introduced into hohlraum modeling by an iterative scheme of coupling the VFP code K-2\cite{Sherlock2017a}  with the radiation hydrodynamics code HYDRA\cite{Marinak2001} via direct thermal conduction coupling for one -dimensional hohlraum modeling, where K-2 supplies the thermal conduction \cite{MarinakAPS}. This direct coupling contrasts with the methods typically employed in MCF SOL simulations, where long timescales are of primary interest. Notably, Zhao \textit{et al.} applied a constant multiplier to all variables of interest:
\begin{gather}
\mathbf{q_e} = c_{eff} \, n \tau_e \frac{T_e}{m_e} \nabla T_e,\quad
\mathbf{R_{T_\parallel}} = -k_{eff} \, n \nabla_\parallel T_e,
\end{gather}
where \(c_{eff}\) and \(k_{eff}\) are the multipliers for the heat flow and thermal force terms, determined from the VFP code \cite{Zhao2017,Zhao2019}. These correction factors were iteratively applied between the VFP and hydrodynamics codes to ensure accuracy. Zhao \textit{et al.} used the VFP code KIPP, which solves the VFP equation directly in one spatial dimension and two velocity dimensions without expansion \cite{Chankin2012, Chankin2014, Chankin_2022}.

Typically, the VFP equations are solved as non-steady-state problems, with thermodynamic profiles evolving over time and providing multipliers for different sets of \(T_e\) and \(n_e\). To address this, Zhao \textit{et al.} applied heating and cooling to the distribution function, directly modifying it to maintain constant temperature and density. However, this method is only applicable to codes that solve for the full distribution function and is challenging to implement robustly in VFP codes that use expansions of the VFP equations, such as the approach employed in this work with the SOL-KiT and previous work using K-2. Nonetheless, the heating and cooling approach was utilized in the HYDRA–K-2 direct coupling \cite{MarinakAPS}.

The coupling methods proposed by Zhao \textit{et al.} \cite{Zhao2017} present particular challenges for dynamic problems such as edge-localized modes (ELMs) and the hohlraum and capsule dynamics in ICF. This is largely because temporal accuracy is not critical in traditional SOL modeling, which primarily assumes steady-state conditions. In contrast, dynamic problems such as ELMs\cite{VASILESKA2021112407} and ICF applications require high temporal accuracy. To address this, Marinak \textit{et al.} \cite{MarinakAPS} employed direct thermal conduction coupling. However, this method incurs high computational expense for achieving numerically stable solutions, and its accuracy in nonlinear regimes remains uncertain. Thus, a novel method improving on previous attempts is required to achieve reliable VFP coupling with hydrodynamic codes in ICF contexts. Importantly, while our focus is on ICF-relevant conditions, the proposed method is broadly applicable to any fusion scheme where kinetic electron effects are significant. In this article, we present a method for coupling VFP and radiation hydrodynamics systems, and we extensively test it for temporal accuracy under ICF-relevant conditions.

The remainder of this article is structured as follows: Section \ref{sec:method} describes the algorithm implemented, \ref{sec:bmark} presents benchmarking of the method and \ref{sec:d+conc} provides a discussion of the findings and conclusions.

\begin{center}
\begin{algorithm}[H]
\SetAlgoLined
\KwResult{System state at time $t_{\text{cycle}} \times Cycle_{max}$}
Initialize system state (profiles $n$, $T$, etc.)\;
Initialize $M \gets$ (initial value or computed from initial fluxes)\;

\For{$Cycle = 1$ \KwTo $Cycle_{max}$}{
    Run Hydro simulation for duration $t_{\text{hydro}}$ (or for $N_{\text{hydro}}$ timesteps) using current $n$, $T$ profiles and parameter $M$\;
    
    Initialize kinetic iteration counter: $k \gets 0$\;
    Compute initial kinetic flux $q^{(0)}_{VFP}$ using the updated $n$, $T$ profiles\;
    \While{$\displaystyle\frac{\left|q^{(k+1)}_{VFP} - q^{(k)}_{VFP}\right|}{q^{(k)}_{VFP}} > tol$}{
         Evolve kinetic simulation for one timestep\;
         Update kinetic iteration counter: $k \gets k+1$\;
         Re-scale physical quantities as required\;
         Update flux: compute $q^{(k+1)}_{VFP}$\;
    }
    Compute $M \gets \frac{q_{VFP}}{q_{SNB}}$\;
}
\caption{VFP-Driven Method}
\label{algo:hkc}
\end{algorithm}
\end{center}

\section{Method}
\label{sec:method}
Here, we present a general method designed for coupling VFP codes with hydrodynamics codes, addressing problems where the classical Fourier heat-law breaks down. The scheme requires relatively minor modifications on both the hydrodynamics and VFP sides. We start by discussing the design logic and requirements for the hydrodynamics codes, followed by modifications to the VFP codes, and conclude with a brief discussion of convergence criteria and efficiency of the method.
\begin{figure}
    \centering
    \includegraphics[width=1.0\linewidth]{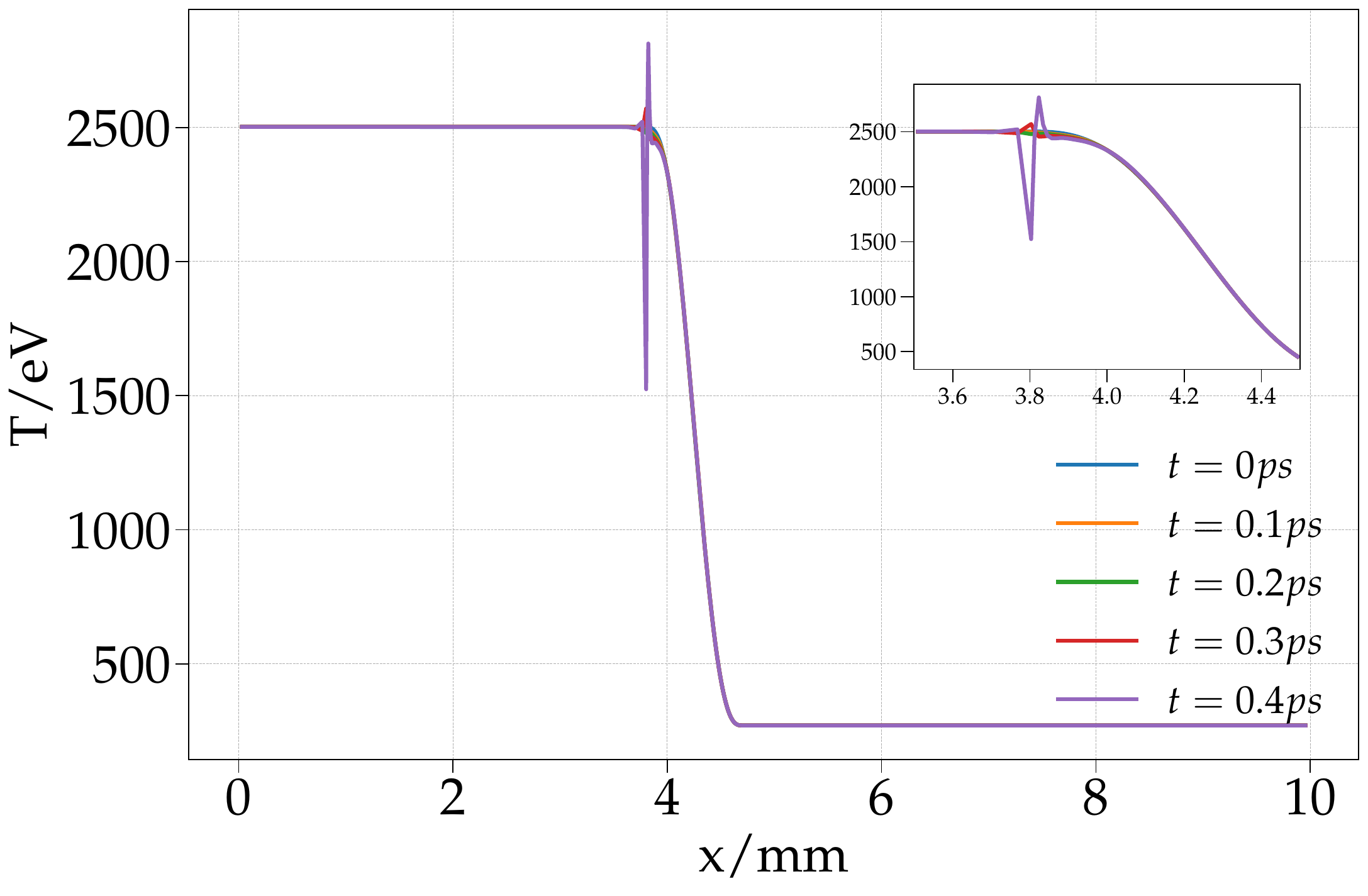}
    \caption{Illustrations of $\nabla\cdot q=const$  coupling techniques and its respective failure modes.}
    \label{fig:couple_divq_fail}
\end{figure}
\begin{figure}
    \centering
    \includegraphics[width=1.0\linewidth]{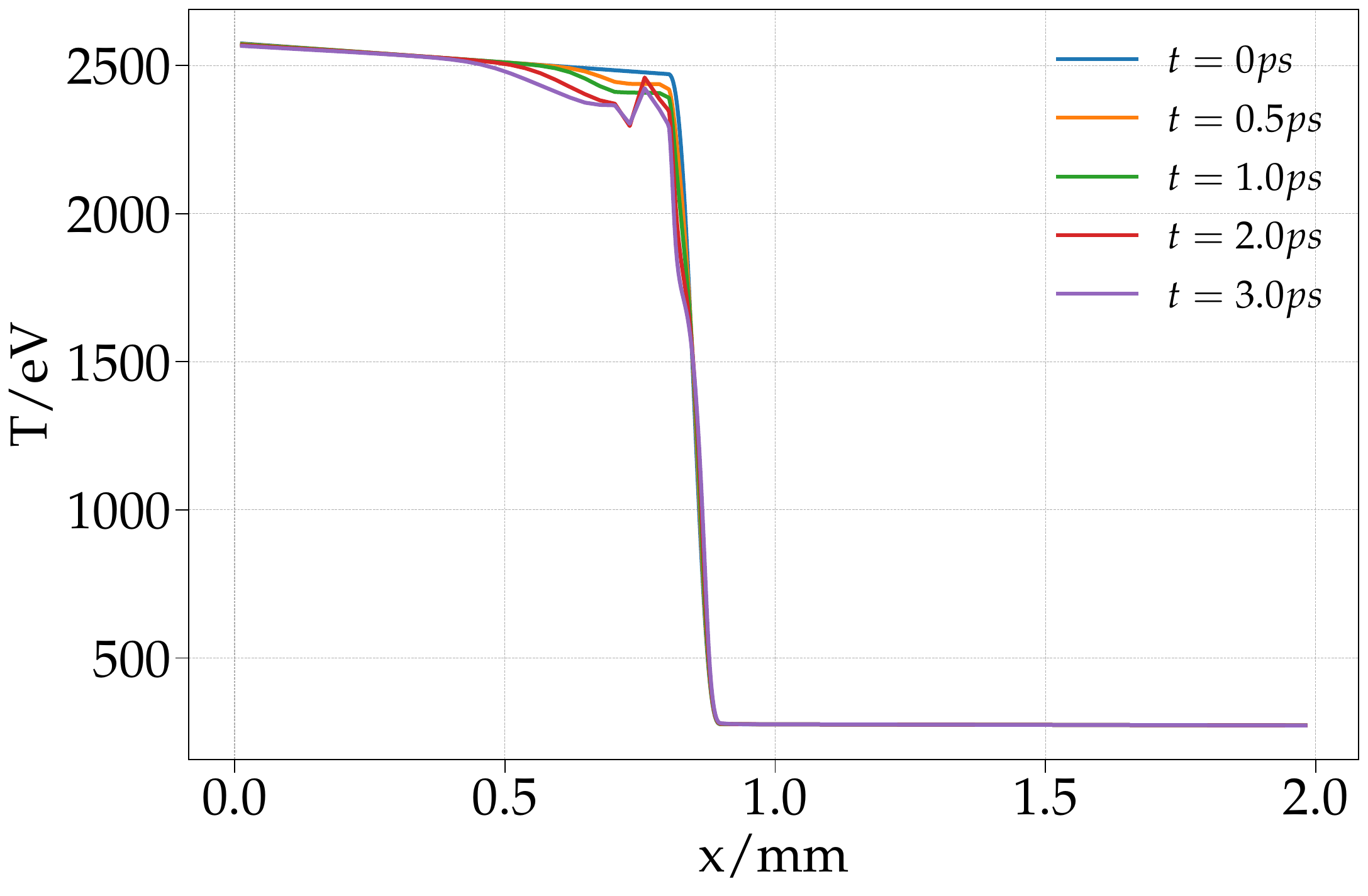}
    \caption{Illustration of Spitzer-Harm multiplier coupling technique and its respective failure modes - formation of a growing numerical artifact.}
    \label{fig:couple_multi_fail}
\end{figure}
The constructed scheme utilizes a similar iterative approach to Zhao et al. \cite{Zhao2017, Zhao2019}, with notable differences to account for the following three conditions:
\begin{itemize}
    \item Ensuring that as many radiation-hydrodynamics codes as possible can utilize the tools developed with only minor modifications.
    \item Minimizing CPU costs for implicit VFP codes.
    \item Maintaining the stability of explicit Rad-Hydro schemes.
\end{itemize}

This method allows multiple hydrodynamic (hydro) time steps to be performed before invoking the more CPU-intensive VFP code to adjust transport properties. Each “cycle” involves running the fluid code up to a specified time  $t_{\text{cycle}}$ , after which the VFP code is executed again. Both the hydro and VFP codes operate with independent time steps($dt_{fluid}$ and $dt_{vfp}$) as necessary for numerical stability. In this work, we refer to one iteration between the hydro and VFP codes as a “cycle.”

Correcting Braginskii’s results, as done by Zhao \textit{et al.} \cite{Zhao2017, Zhao2019} in the ICF context, is difficult because the extreme conditions involve large heat flows. Therefore, large correction factors during peak laser intensity are expected. These correction factors could lead to numerical instabilities if large time steps are used illustrated in Fig. \ref{fig:couple_multi_fail}.  Consequently, this makes adjusting Braginskii heat-fluxes less feasible. Additionally, pre-heat effects where heatflow occurs ahead of the bulk temperature gradient in a region with no or negligible $\nabla T$ cannot be simply accounted for via a multiplier, making this approach challenging to implement. 

Single-step correction methods, such as the direct coupling method (i.e., utilizing $\nabla \cdot q_{VFP}$ from the converged VFP equations) used by Marinak \textit{et al.}, are challenging to use due to the high CPU time required to converge VFP codes. For systems such as hohlraums, convergence time can exceed 60 minutes for a single VFP iteration. Thus, using a direct coupling method —particularly with implicit VFP codes like SOL-KiT \cite{Mijin2021comp} —is computationally expensive. Attempting to use a clamped valued of $\nabla\cdot\mathbf{q}_{_\mathrm{VFP}}$ over a significant number of hydro-code time steps would reduce numerical accuracy and even introduce spurious artefacts (e.g. maxima \& minima) since in practice the divergence of the heat-flux changes rapidly in a dynamic system. An illustration of this potential issue is highlighted in Fig. \ref{fig:couple_divq_fail}. Therefore, a multiplier approach is more advantageous, as the underlying heat-flow model will follow natural gradient developments. However, it should be applied to an underlying transport model where the multipliers are not large, and pre-heat is accounted for.

Thus, the setup for this work utilizes a multiplier correction term on the SNB heat flow and we call it the "VFP-Driven" method. This approach is based on the fact that the SNB model in certain contexts has a smaller deviation than Spitzer from the VFP equation, as shown in \cite{Brodrick2017, Sherlock2017a}. Hence, the SNB model will be stable for ICF problems, allowing for multiple fluid steps with a constant correction factor—within some error margin—depending on the iteration time. Importantly, the method prevents  multipliers from being negative by enforcing it to be 1 in those instances. Inaccuracies in the SNB heat-flow value means that regions where the VFP and SNB predicted heat flows are in opposite directions (though typically small in magnitude) often exist resulting in negative multiplier. This restriction promotes time-step stability without drastically altering behavior due to these regions having heat-flows with small magnitudes.

Hence, the temperature equation solved in this work is:
\begin{equation}
C_v \frac{\mathrm{d}T(x,t)}{\mathrm{d}t} = \nabla \cdot \left( \frac{q_{VFP}}{q_{SNB}} \mathbf{q_{SNB}}(x,t) \right) + S,
\end{equation}
where $C_v$ is the specific heat capacity, $S$ represents all other source terms, and $q_{SNB}$ is the heat flow calculated using the SNB model \cite{Schurtz2000}. Furthermore, the problem encountered by the VFP code SOL-KiT is limited by calculating the plasma parameter $n \lambda_{db}^3$. We enforce that the plasma should be as close to ideal as possible ($1/n\lambda_{db}^3 \leq A$), where $A$ is typically set to a value less than 5. SOL-KiT does not model regions where $A$ exceeds this threshold, and standard SNB transport is used in these areas. This approach improves computational time in SOL-KiT because regions of high density are typically prohibitively expensive by standard numerical algorithms to solve the VFP equation. Thus, for hohlraums the shocked cold hohlraum wall cannot be modeled kinetically. However, this is acceptable as the scale lengths deep into the plasma are small, leading to distribution functions close to a local solution. Nevertheless, this may not hold for all scenarios. The designed algorithm is outlined in Algorithm \ref{algo:hkc}.

As done by Zhao \textit{et al.} \cite{Zhao2019a} and Marinak \textit{et al.}\cite{MarinakAPS}, the VFP code must retain the thermodynamic profiles while allowing the distribution function to change termed as "Re-scale" in Algo. \ref{algo:hkc}. For cases involving the full VFP equation, the methods proposed by Zhao \textit{et al.} can be applied directly . However, for codes using expansions of the VFP equation (such as IMPACT \cite{Kingham2002}, K-2 \cite{Sherlock2017a}, KALOS \cite{Bell2006}, IMPACTA \cite{Thomas2012}, and SOL-KiT \cite{Mijin2021comp}), a novel approach is required as discussed previously. For these expanded methods, maintaining the zeroth-order properties of the distribution function $f_0$ keeps the thermodynamics constant while allowing the distribution function to evolve—referred to as ``\textit{rescaling $f_0$}'' and ``\textit{maintaing $f_0$}''in this work. This approach ensures that the fluxes correspond to static thermodynamic profiles.

To achieve this, $f_0$ must be rescaled at every VFP time step to prevent time evolution of $T_e$ or $n_e$. Let $f_0$ represent the $l=0$ electron harmonic at the end of the fluid step. Thus, we know:
\begin{subequations}
\begin{gather}
\int_0^\infty v^2 f_0 \mathrm{d}v = n(x), \
\frac{m_e}{2}\int_0^\infty v^4 f_0 \mathrm{d}v = E(x) = \frac{3}{2}nT,
\end{gather}
\end{subequations}
where the quantities on the RHS are state variables at the end of the fluid step. Let us define a new distribution function $f_0’(v)$ with particle and energy moments $n’$ and $E’$, respectively. The new distribution function is:
\begin{equation}
f_0’(v) = af_0(bv).
\end{equation}
We now have all the information required to calculate $a$ and $b$ to rescale the old distribution function. Given:
\begin{equation}
\int_0^\infty v^2 af_0(bv) \mathrm{d}v = n’,
\end{equation}
and performing a change of variables (e.g., $v’ = bv$) results in:
\begin{subequations}
\begin{gather}
\int_0^\infty \frac{a}{b^3} v’^2 f_0(v’) \mathrm{d}v’ = n’,\
\frac{a}{b^3}n = n’.
\end{gather}
\end{subequations}
Following the same procedure for the energy moment, we derive the following definitions for $a$ and $b$:
\begin{subequations}
\begin{gather}
a = \frac{n’b^3}{n} \; ,\\
b^2 = \frac{n’E}{nE’}\;.
\end{gather}
\end{subequations}
Some care is required when applying this to a discrete grid, as the scaled $v$ values may land between the discrete velocity grid points. Thus, the analytical form of $b$ informs us of the nearest two velocity cells, $m$ and $m+1$, between which the analytical value $vb$ lies. The numerical version of $f_0’(v_n)$ is then:
\begin{equation}
f_0’(v_n) = c \left( d f_0(v_{m_n}) + (1 - d) f_0(v_{m_{n+1}}) \right),
\end{equation}
resulting in the following scaling conditions for $c$ and $d$:
\begin{subequations}
\begin{gather}
4\pi \sum_n c \left( d f_0(v_{m_n}) + (1 - d) f_0(v_{m_{n+1}}) \right) v_n^2 \Delta v_n = n’, \\
4\pi \sum_n c \left( d f_0(v_{m_n}) + (1 - d) f_0(v_{m_{n+1}}) \right) v_n^4 \Delta v_n = E’,
\end{gather}
\label{eq:maintain}
\end{subequations}
which can be solved to obtain $c$ and $d$. The pathological case of $b = 1$ must be handled separately, where the solution is simply $f_0’ = af_0(v)$. Note that spatial notation is omitted for clarity. To maintain the profile, let $f_0' = f_0^{n+1}$ and $f_0 = f_0^{n}$, thus rescaling every VFP time step allows us to retain the thermodynamic quantities.

Thus, the VFP-driven method requires the following two minor changes 
\begin{itemize}
    \item The Rad-Hydro code needs to have a hook for a multiplier for its heat-flux
    \item The VFP codes needs to have a hook for modifying the distribution function according to Eq. \ref{eq:maintain}.
\end{itemize}
The largest change required will be to the VFP codes as this is not usually available, however, is available in the code used in this paper SOL-KiT\cite{Mijin2020Thesis}.

\subsection{Convergence and Efficiency}

To achieve consistent repeatability and efficiency, we run the VFP code until a convergence criterion is met. In this work, we use a relative difference convergence metric defined as:
\begin{equation}
conv = \max\left( \frac{|q^{n+1}_{\mathrm{VFP}} - q^n_{\mathrm{VFP}}|}{|q^n_{\mathrm{VFP}}|} \right),
\end{equation}
where  $q_{\mathrm{VFP}}^{n+1}$  is the heat-flow output after  N  time steps, each of duration  $dt_{\mathrm{vfp}}$ , and  $q_{\mathrm{VFP}}^{n}$  is the heat-flow output from the previous check. We exclude small relative differences less than  $1 \times 10^{-16}$  and cases where  $q_{\mathrm{VFP}}^{n} \sim 0$ , which could result in undefined or infinite values (NaN/infs) in the metric.

Previous studies \cite{Kessler2010, Kessler2011, Zhao2017} suggest monitoring the convergence of harmonics  $f_{l>0}$ . However, we find that focusing on the convergence of heat flow is more relevant for our purposes. The convergence of  $f_{l>0}$  does not always correspond to significant changes in heat flow. Since our main objective is to obtain accurate heat-flow values, monitoring the effective change in calculated heat flow provides a more meaningful convergence metric. This approach can also reduce computation time, depending on the required accuracy.

In this work, we set the convergence criterion to  $\text{conv} < 0.02$  with  N = 360 . This means the calculation is checked every  $360 \times dt_{\mathrm{vfp}}$ , where  $dt_{\mathrm{vfp}}$  varies between  0.03  and  18\text{fs}.

To accelerate the VFP calculation at the end of each cycle, we initialize the VFP code differently from the standard approach. Typically, a VFP code starts with a Maxwellian distribution  $f_0$  based on the fluid density and temperature $(n,T)$ and sets $f_{l>0}$  to zero, including  $f_1 = 0$. In our coupled method, however, we use the converged  $l = 1$  harmonic  $f^{n-1}_1$  from the previous cycle’s electron distribution function, along with an updated Maxwellian  $f^{n}_0$  reflecting the latest fluid  $(n^{n}, T^{n})$ . The higher-order harmonics  $f^{n}_l$  for  $l > 1$  are still initially set to zero. This approach is similar to the one used by Kessler \textit{et al.}, who incorporated prior information to update particle velocities in their MC-fluid coupling \cite{Kessler2010, Kessler2011}. We have found that this method enhances computational efficiency, especially when the thermal characteristics of the system change minimally between cycles. While it’s only applicable when cycle-to-cycle changes are small, this strategy can reduce the total CPU time per cycle by up to 50\%.

\section{Benchmarking}
\label{sec:bmark}
\begin{table*}
\centering
\begin{tabular}{@{}llllll@{}}
\toprule
Benchmark & Unit Tested & Grid Resolution  & L(\si{cm}) & dt(\si{ns})& Cycles\\ 
\midrule
Epperlein-Short & Re-scaling  & $63$ & Variable & $2\times10^{3}$& n/a\\
Tanh Ramp & Integrated & $100$ & $0.5$ & $10$ & $5$\\
Hohlraum Test & Integrated & $123$ & $0.4$ & $10^{-2}$ &$5$\\
\bottomrule
\end{tabular}
\caption{Validation benchmarking of VFP-Driven method}
\label{table:vfp_driven_benchmarking}
\end{table*}
The aim of the VFP-Driven method is to provide a general tool for modeling electron kinetics in ICF plasmas. This requires the “maintain” method to preserve the thermodynamic profile for ICF-like plasmas, characterized by large thermal density gradients, variation in material and average ionisation states $\overline{Z}$. Additionally, it must be able to replicate known non-local transport phenomena such as the reduction in decay rate of short wavelength thermal pertubations (e.g. Epperlein and Short \cite{Epperlein1991}). Furthermore, the method is expected to relax the thermal profile of hohlraums without causing stability issues and to be more accurate than alternative transport models, provided there are no source terms. The tests conducted to validate this method are summarized in Table \ref{table:vfp_driven_benchmarking}.

\subsubsection{Epperlein-Short test}
\pgfplotsset{
    width=12.5cm,
    height=9cm,
    every x tick label/.append style = {font=\large},
    every y tick label/.append style = {font=\large},
}
\begin{figure}
    \resizebox{0.8\columnwidth}{!}{
   \begin{tikzpicture}
    \begin{loglogaxis}[
        xlabel = \(k \lambda_{ei}^B\),
        ylabel = {\(\kappa / \kappa^B\)},
        cycle list name=color list,
        ytick={0.1,1.0},
        legend style={draw=none},
        mark options={scale=2},
        ylabel style={yshift=-0.5em},
        ylabel style={font=\large},
        xlabel style={font=\large},
        grid=both,
        grid style={line width=.2pt, draw=gray!10}
    ]
    \addplot [
        domain=0.0075:2, 
        samples=100, 
    ]
    {1 / (1 + 1 /((1 / (51.409789 * x*x)) + (1 / (4.4682314 * x))))};
    \addlegendentry{SOL-KiT Fit\cite{Mijin2021comp}}
    \addplot[
    mark=square,
    only marks
    ]
    coordinates {
    (0.0075, 0.995127)(0.02, 0.987376)(0.04, 0.963115)(0.075, 0.90211)(0.2, 0.68982)(0.4, 0.483059)(0.75, 0.309565)(2.0, 0.128354)
    };
    \addlegendentry{SOL-KiT $l = 1$}
    \addplot[
    mark=x,
    only marks
    ]
    coordinates {
    (0.0075, 0.995)(0.02, 0.985)(0.04, 0.956)(0.075, 0.886)(0.2, 0.662)(0.4, 0.465)(0.75, 0.314)(2.0, 0.153)
    };
    \addlegendentry{Maintain $l = 1$}

    \end{loglogaxis}
    \end{tikzpicture}
    }
    \caption{Epperlein-Short test for the maintain method. Standard SOL-KiT results are included from \cite{Mijin2021comp}. Easily noted that the maintain method is able to get within some margin of error to the expected $\kappa / \kappa^{B}$ for all values of $k\lambda^B_{ei}$ except at $k\lambda^B_{ei} = 2$, this can be explained by not having converged fully. The fit is based on the highest harmonic result from the \cite{Mijin2021comp}.}
    \label{fig:EpperleinShortTest}
\end{figure}
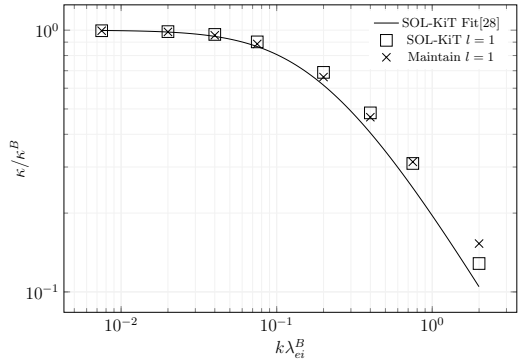

The Epperlein-Short test is a classic non-local test used to evaluate the capability of non-local transport codes to handle increasingly collisionless problems. Although it is not an ideal test for assessing the accuracy of a model in practical applications—many models, such as the SNB model, have passed the Epperlein-Short test but performed poorly in real-world scenarios \cite{Brodrick2017}—failure to achieve high accuracy in this test suggests that the model will likely struggle in practical applications. It is well known that the decay rate $\gamma$ is related to the conductivity ratio by the expression $\gamma / \gamma_B = \kappa / \kappa_B$, where $\gamma_B$ and $\kappa_B$ are the Braginskii thermal decay rate and conductivity, respectively\cite{Brodrick2017}. Hence, the Epperlein-Short test primarily evaluates the system’s ability to find accurate multipliers, making it an ideal test for assessing the maintain method’s capability to determine accurate correction factors.

Since the analysis focuses on hydrogen ($Z = 1$), SOL-KiT is \textbf{not} employed in the Lorentz limit for this specific test. As seen in Fig. \ref{fig:EpperleinShortTest}, when compared directly with the SOL-KiT findings from \cite{Mijin2021comp}, the maintain method successfully finds accurate multipliers. A slight discrepancy at $k\lambda_{ei}^B = 2$ is observed, which can be attributed to convergence errors in both the SOL-KiT result and the maintain method. Nonetheless, the high accuracy in finding the relevant multipliers provides reasonable confidence in the maintain method’s ability to accurately represent heat flows, given a static profile.

\subsubsection{Thermal decay of Tanh ramp}
\label{ch:2_sec:tanh}
\begin{figure*}
    \centering
    \includegraphics[width=\columnwidth]{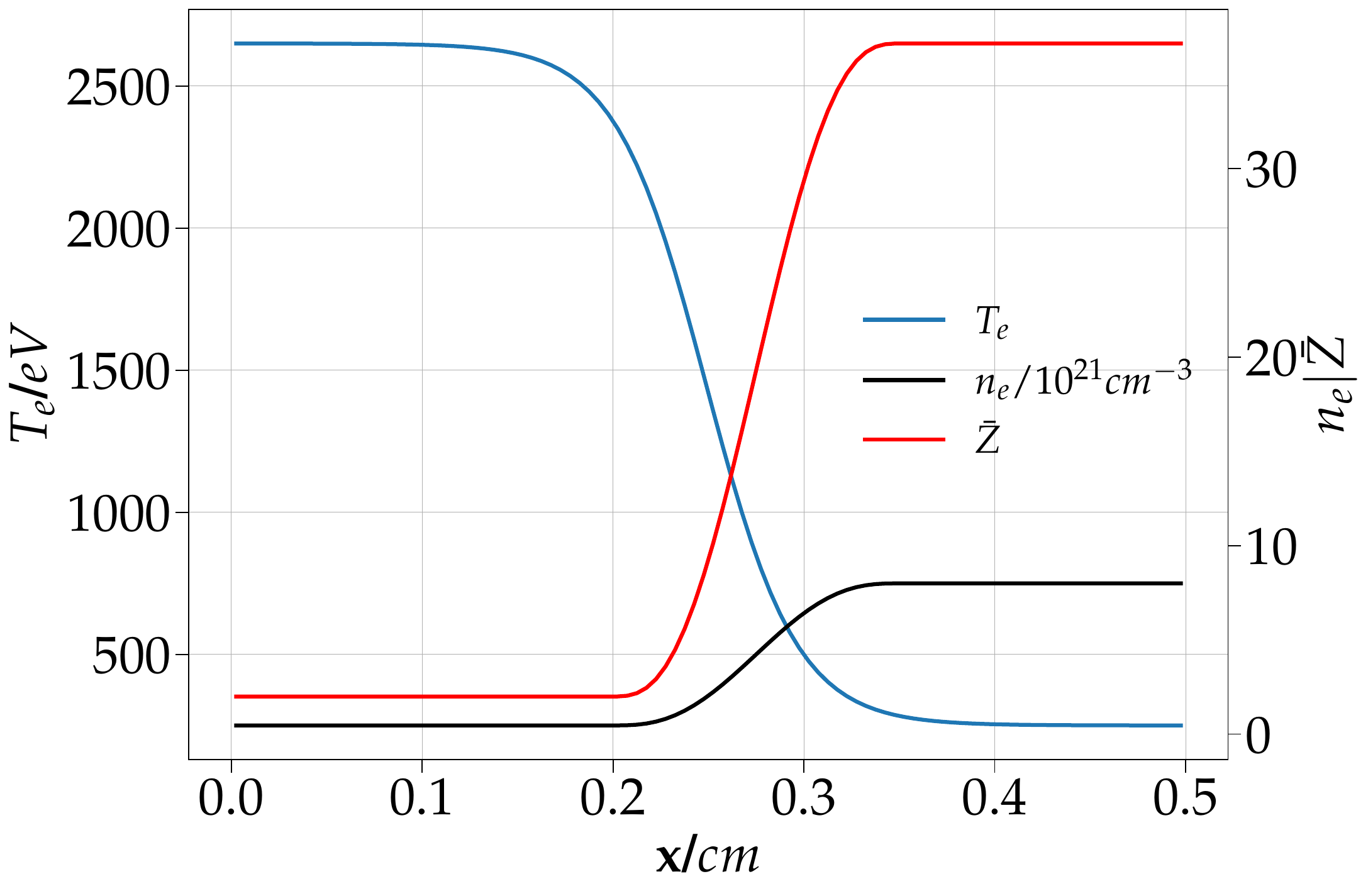}\hfill
    \includegraphics[width=\columnwidth]{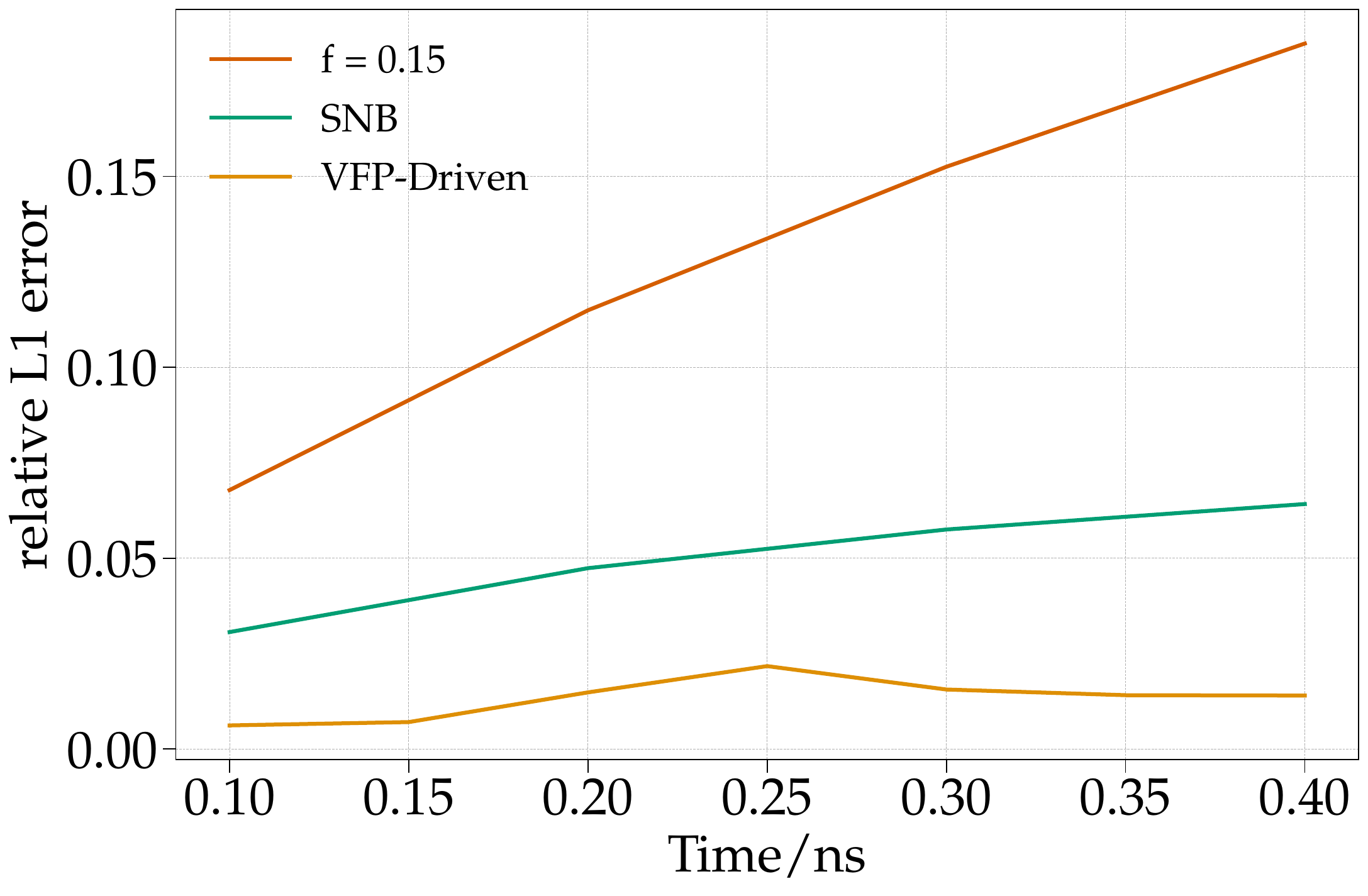}
    \caption{Smooth Tanh ramp thermal decay test. Left) Initial conditions for the test problem; ionisation remains constant, and the system is run as a single temperature system. Right) relative L1 error  of three different models, flux-limited Spitzer with $f=0.15$, SNB and VFP-Driven Method. Note that the error growth in the VFP-Driven method is inconsistent and slow compared with SNB \& $f = 0.15$, which both grow roughly linearly over time.}
    \label{fig:tanhtestinit}
\end{figure*}

A smooth ramp is used to test the capabilities of the VFP-Driven method. This test is designed to evaluate the VFP-Driven method’s accuracy in smooth problems. We initiate a $\tanh$ ramp over a distance of $0.5 \si{cm}$, with a temperature gradient from $2.7 \si{keV}$ to $0.27 \si{keV}$, a density from $0.4\times10^{21} \si{cm^{-3}}$ to $8\times10^{21} \si{cm^{-3}}$, and a mean ionization from 2 to 36. The system is modeled as an ideal gas, with the mean ionization and number density fixed. The test is conducted as a pure thermal decay problem driven by electron thermal conduction. These constraints ensure a like-to-like comparison between SOL-KiT (in pure VFP mode) and the VFP-Driven method.

We utilise a relative L1 error \cite{vanderholst_2011} in the $T_e$ profile as a measure of overall accuracy
\begin{equation}
    E_{L1}= \frac{\sum_{x=1}^{n_x}|T_x^{\, true} - T_x^{\, sim}|}{\sum_{x=1}^{n_x}|T_x^{\, true}|},
    \label{eq:are}
\end{equation}
where $T_n^{, true}$ is the temperature obtained from a pure kinetic calculation, i.e., running SOL-KiT independently and is summed over the entire grid $n_x$. As shown in Fig. \ref{fig:tanhtestinit}, the overall accuracy of the SNB and VFP-Driven methods is better than that of the Spitzer model. This improvement is partly due to the flux-limiter in the Spitzer model suppressing heat flow too much at peak temperatures, which the SNB and VFP-Driven methods capture more accurately. The VFP-Driven method demonstrates greater accuracy than the SNB model and, importantly, has slow and inconsistent error growth over time, which is critical for temporal accuracy.

\subsubsection{Thermal decay of hohlraum}
\begin{figure*}
\centering
    \includegraphics[width=\columnwidth]{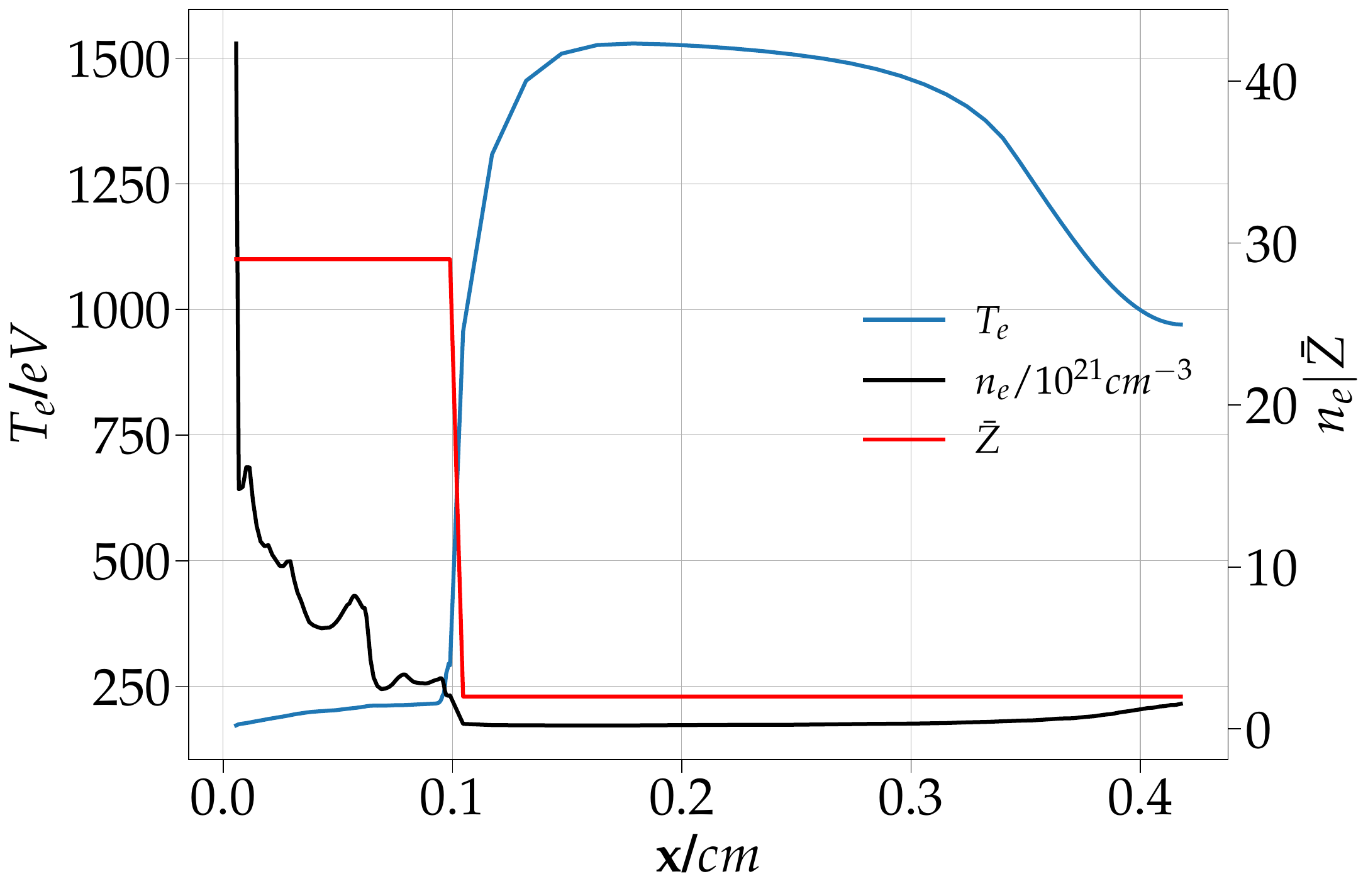}\hfill
    \includegraphics[width=\columnwidth]{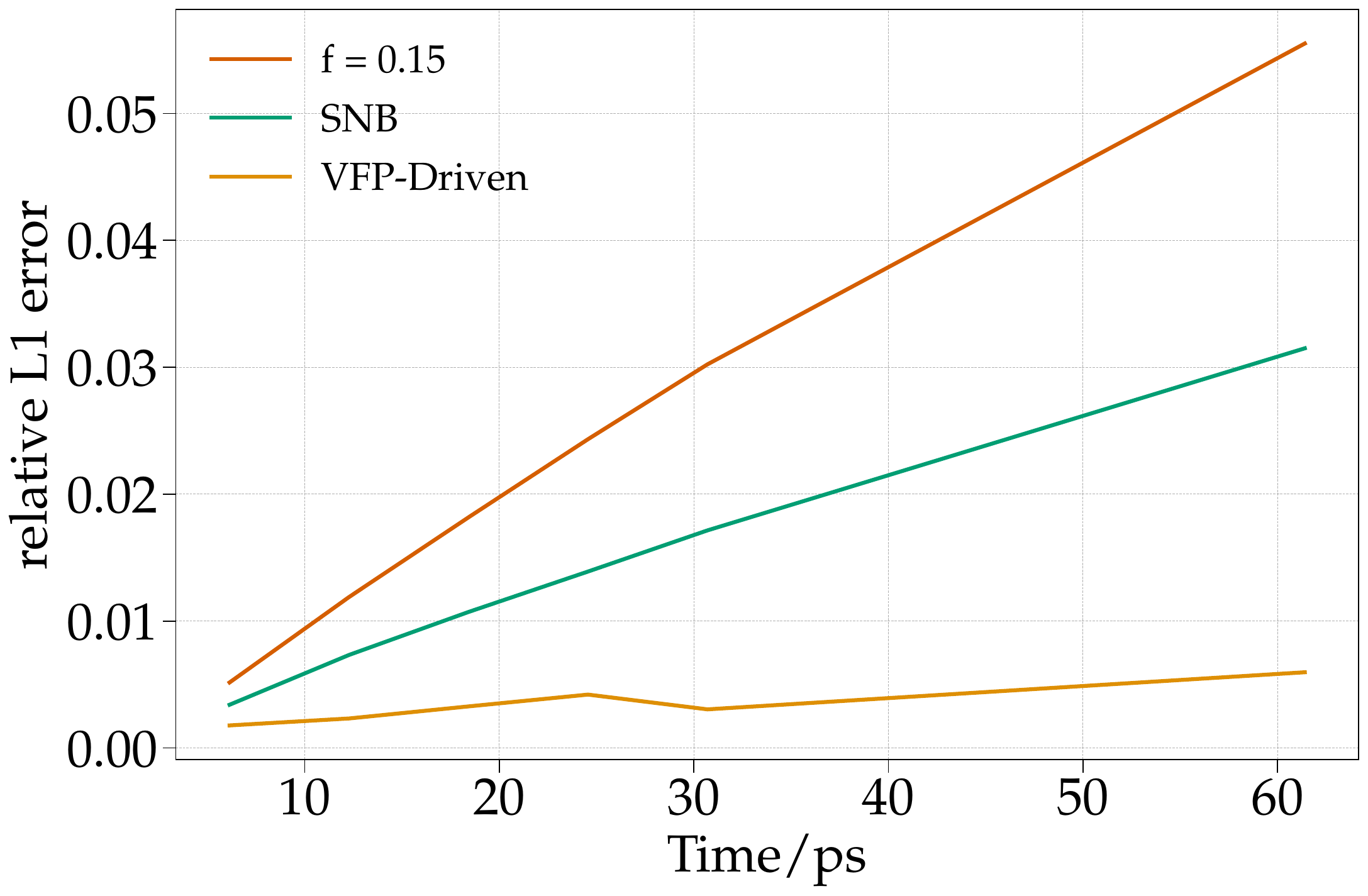}
    \caption{Hohlraum thermal decay test. Left: Initial conditions for the test problem. Ionization remains constant, and the system is run as a single temperature system. Right) relative L1 error for three different models: flux-limited Spitzer with $f=0.15$, SNB, and the VFP-Driven method. Note that error growth in the VFP-Driven method is inconsistent and slow, while both SNB and $f=0.15$ grow approximately linearly over time.}
\label{fig:hohltestinit}
\end{figure*}

To ensure the feasibility of running hohlraum conditions, we evaluate the accuracy of the VFP-Driven method using a thermally decaying hohlraum. The initial conditions are based on a high gas-fill copper hohlraum without dense matter. No other sources are present except for thermal conduction and $\mathrm{PdV}$ work, i.e., the system is run as a single-source problem. The system is assumed to be fully ionized and modeled as an ideal gas. Although this setup does not represent a realistic hohlraum scenario (due to the assumption of ideal gas behavior and fixed ionization), it allows for a fair comparison between kinetic models and the VFP-Driven method.

The initial conditions and the propagation of relative L1 error (Eq. \ref{eq:are}) over time are shown in Fig. \ref{fig:hohltestinit}. After $60 \si{ps}$, the VFP-Driven method exhibits the lowest relative L1 error. Additionally, the error growth in time is approximately linear for the published methods, while the VFP-Driven method maintains inconsistent error growth, as seen previously. This demonstrates several key findings: the VFP-Driven method is more accurate than current alternatives for hohlraum-like conditions, and it can stably decay steep hohlraum gradients without causing numerical instabilities or significant errors. These results provide sufficient evidence that the VFP coupling methodology can be applied to full-scale hohlraum simulations.

\section{Conclusion}
\label{sec:d+conc}

We have devised a novel method for coupling VFP to plasma hydrodynamic equations, creating a general and versatile approach that can be integrated with a wide range of hydrodynamic codes. This approach, termed VFP-Driven Hydrodynamics, employs an external VFP code to drive electron thermal conduction in the radiation-hydrodynamics code through adjustments to a native thermal conduction model using a dynamic multiplier. For highly dynamic problems such as in ICF , it performs optimally when paired with the SNB conduction model rather than Spitzer-Harm. Testing on classical thermal decay problems shows that the method matches the accuracy of standalone VFP calculations even as nonlinearity increases. Additionally, in ICF-relevant scenarios, the method has demonstrated greater accuracy than established models while maintaining stability under hohlraum-specific conditions. Moreover, the SNB coupling method could be extended to other highly dynamic situations where non-local transport or indeed any other kinetic effects may be of importance, such as ELM simulations typically handled via PIC \cite{VASILESKA2021112407}

Although promising, this method has so far only been validated in 1D studies. Future studies will be essential to confirm if the improved accuracy seen in 1D carries over to 2D and 3D cases, where unique challenges in high-dimensional spaces may present themselves. Additionally, future work could also focus on optimizing the computational efficiency of VFP-Driven Hydrodynamics, particularly as studies expand to 2D and 3D models

Further validation is also needed to assess applicability in the presence of magnetic fields. We propose that this method could be extended relatively straightforwardly to drive other transport relations, such as Ohm’s law, and thereby provide corrections to magnetic induction terms, particularly Biermann, Nernst, and Advection, which are currently addressed via multipliers \cite{sherlock2020,Campbell2020,Brodrick2018}.

The approach outlined in VFP-Driven Hydrodynamics may also serve as a valuable framework for coupling kinetic and hydrodynamic models across various plasma applications, offering a foundation for future advancements in transport modeling.

\begin{acknowledgments}
We wish to acknowledge the support of the Centre for Inerital Fusion Studies at Imperial College London for funding this research. We acknowledge computational resources and support provided by the Imperial College Research Computing Service (http://doi.org/10.14469/hpc/2232)"
\end{acknowledgments}

\section{Data Availability Statement}
Data available on request from the authors

\section{Author Contributions}
\textbf{Abetharan Antony}: Conceptualization, Investigation, Visualization, Methodology, Formal Analysis, Software, Writing - Original Draft. \textbf{Stefan Mijin}: Software, Methodology, Writing - Review \& Editing. \textbf{Robert Kingham}: Conceptualization, Supervision, Writing - Review \& Editing. \textbf{Marty Marinak}:  Conceptualization, Writing - Review \& Editing. 
\bibliography{library}

\end{document}